\documentclass[11pt]{article}

\usepackage[utf8]{inputenc}
\usepackage[T1]{fontenc}
\usepackage{lmodern}
\usepackage[margin=1in]{geometry}
\usepackage{graphicx}
\usepackage{booktabs}
\usepackage{multirow}
\usepackage{amsmath,amssymb}
\usepackage{authblk}
\usepackage{caption}
\usepackage{xcolor}
\usepackage{mdframed}
\usepackage[numbers,sort&compress]{natbib}
\usepackage[colorlinks=true,allcolors=blue]{hyperref}

\usepackage{csquotes}

\newenvironment{significance}{%
  \begin{mdframed}[backgroundcolor=gray!8,linecolor=gray!40,roundcorner=4pt,
                   innertopmargin=8pt,innerbottommargin=8pt]%
  \noindent\textbf{Significance Statement}\par\smallskip\small}{%
  \end{mdframed}\medskip}

\title{\bfseries Free-form Association Tasks Reveal Stereotype Hallucination in Large Language Models}

\author[a]{Xinrui Chloe Zhao}
\author[a]{Douglas Guilbeault}
\author[a]{Amir Goldberg}
\affil[a]{Stanford Graduate School of Business, Stanford University, Stanford, CA 94305}

\date{\today}

\begin{document}

\maketitle

\begin{abstract}
\noindent
Recent studies argue that LLMs can predict human stereotypical judgments. Yet whether LLMs emulate the cognitive processes underlying human stereotypes, or merely retrieve learned associations to solve prediction tasks, remains unclear. Prior work examines LLMs' stereotypes in either (i) controlled judgment tasks like multiple choice surveys, or (ii) contexts constrained by conventionalized and predictable group biases. Here, we compare the structure of the stereotypes that humans and LLMs exhibit in the interpretation of free-form stimuli, namely abstract art and Rorschach blots, which lack pre-established cultural meanings. We recruit participants across five social domains (gender, partisanship, personality, urbanicity, and lifestyle) and elicit both first-order (direct personal interpretations) and second-order responses (predictions about how members of social groups will interpret the stimuli); we replicate this design with two multimodal models (GPT-4o mini and Llama-3.2-11B-Vision-Instruct). Humans and LLMs differ not only in magnitude but in the qualitative nature of their stereotypes. Human first-order responses display heterogeneity with minimal group structure. When predicting group responses, humans engage in ``stereotype exaggeration'' by moderately amplifying first-order tendencies while preserving diversity. By contrast, LLMs exhibit homogeneous first-order responses, and yet generate stark second-order stereotypes that neither amplify existing first-order tendencies nor reflect actual human group differences, a process we term ``stereotype hallucination.'' LLMs continued to hallucinate stereotypes even when fine-tuned on the response data of actual participants. These findings suggest significant limitations in the use of LLMs to model and predict human behavior in novel contexts involving diverse interpretations.
\end{abstract}


\medskip

\begin{significance}
Prior work shows that stereotypes about ``how different social groups see things'' emerge not from actual differences in perception but from the tendency to construct simplified narratives when reasoning about others (i.e., from our second-order beliefs about what others believe). Here, we evaluate whether LLMs capture the effects of second-order beliefs on stereotyping in novel interpretative environments. Comparing how humans and LLMs interpret arbitrary, free-form stimuli, we show that second-order beliefs induce stereotype exaggeration in humans (reifying ground-truth biases), whereas in LLMs they induce stereotype hallucination (distorting and untethering interpretation from human ground truth). Second-order beliefs produce qualitatively different stereotyping in humans and LLMs, revealing significant limits on LLMs' ability to model and predict human interpretation in ambiguous, novel contexts.
\end{significance}

\section{Introduction}
For decades, social science has viewed stereotypes as stemming from a process of statistical exaggeration, whereby observed attributes of people are mistakenly viewed as broadly predictive of qualities in other people belonging to the same social group. Indeed, in his landmark treatise \textit{The Nature of Prejudice}, Allport explicitly defines stereotypes as ``exaggerated belief[s] associated with a [social] category'' (p. 191) \cite{allport1954}. On this account, stereotypes are the byproduct of the cognitive need to simplify perceptions of human variation by categorizing people using statistical generalizations based on their group membership \cite{taylor2015categorization, bai2020diversity, guilbeault2024online}. This bias is readily on display in studies that examine people's second-order beliefs, i.e., their beliefs about what people in particular social groups believe. A number of studies show that second-order beliefs induce exaggerated perceptions of group similarities that underestimate heterogeneity in people's beliefs, regardless of whether its with respect to one's ingroup or outgroup \cite{mize2024divergence, guilbeault2024exposure, van2024imagined}. For example, a recent paper showed not only that Republicans and Democrats significantly underestimated variation in each other's political views, but also underestimated variation within their own parties \cite{dias2025american}. Learning how to identify and mitigate stereotypes in second-order beliefs remains a longstanding goal across the social sciences, from psychology to sociology.

Recent advances in artificial intelligence suggest that large language models (LLMs) may provide an especially promising approach to predicting, simulating, and even intervening on social biases in human judgments. Numerous studies show that LLMs reproduce many common stereotypes in their responses across a range of environments \cite{caliskan2017semantics, kozlowski2019geometry, charlesworth2022historical, charlesworth2024extracting, guilbeault2024online, guilbeault2025age, zewail2026moral}. This has led numerous studies to argue that LLMs can predict human biases and may be able to minimize them through tailored interactions \cite{leidinger2024llms, boissin2025dialogues, costello2024durably}. Indeed, several studies show that access to participants' demographic data allows LLMs to reliably predict participants' responses in online surveys and behavioral experiments \cite{park2022social, park2023generative, chen2023emergence,  park2024generative, hewitt2024predicting, kozlowski2025simulating}. Recent work goes so far as to argue that LLMs trained on historical data can predict how people belonging to different social groups will respond to future events even when these events are not present in the LLMs' training data, such as a recent paper by Kozlowksi et al. (2026) which argues that LLMs trained on internet data from before the COVID-19 pandemic can accurately predict how real Republicans and Democrats responded to COVID-19 vaccine mask mandates on Twitter \cite{kozlowski2024silico}. These advances have led many to argue that LLMs can serve as a stand-in model organism for the social sciences, given their purported ability to reproduce human social cognition \textit{in silico} \cite{filippas2024large, park2022social, park2023generative,  park2024generative, piantadosi2024concepts, ananthaswamy2024close}.

However, in parallel, a growing body of work provides evidence of foundational differences between human and LLM cognition, suggesting limitations on the viability of LLMs as cognitive models of stereotypes \cite{stella2023using}. These include differences spanning several key functions of the mind, from perception \cite{nadler2023divergences, bowers2023deep} to the interpretation of categories in everyday language \cite{nadler2025statistical, palmarini2024abstract, beger2025ai, mitchell2023debate}. An alternative perspective is gaining traction that LLMs are better viewed as a distinct form of intelligence with cognitive and social foundations that vary qualitatively from humans, rather than as tools capable of reproducing and replacing human subjects in social psychological research \cite{brinkmann2023machine, farrell2025large}. Recent work goes further in arguing that the data LLMs generate for social science is systematically biased and unrealistic, due in part to its inability to capture the underlying heterogeneity and complexity of human behavioral data \cite{xie_evaluating_2026}. It thus remains unclear whether LLMs representations emulate the structure of human stereotypes -- licensing their use as psychological model organisms -- or whether they merely rely on learned associations in training data to predict behavior through potentially divergent computational processes.

Past research on comparing humans and LLMs has been limited in its ability to evaluate whether successful predictions of human stereotypes stem from similar or distinct computational processes in part because they have primarily focused on predicting human responses in: (i) highly controlled online surveys and experiments associated with large amounts of relevant training data \cite{park2024generative, hewitt2024predicting, kozlowski2025simulating}, some of which may suffer from systematic data leakage between training and testing data \cite{balloccu2024leak, kaneko2025investigating}; and (ii) real-world contexts in which human responses are shaped by highly entrenched and institutionalized group boundaries, such as partisan responses to online news \cite{kozlowski2024silico, kozlowski2025simulating}, for which human responses are expected to be particularly constrained and predictable. An important, underexplored question concerns the extent to which LLMs can predict and reproduce human stereotyping in novel contexts that are less constrained by entrenched cultural meanings and which go beyond LLMs' training data.

To address this limitation, we compare how humans and LLMs express stereotypes in the interpretation of free-form stimuli, namely abstract art and Rorschach blots, which lack pre-established cultural meanings. Recent work shows that despite inducing considerable variation in human responses, free-form stimuli also expose cognitive constraints and biases that underlie human interpretations \cite{davis2019does, guilbeault2021experimental}. For this reason, free-form stimuli provide an ideal environment for examining first, whether LLMs reproduce variation in human responses, and second, whether humans and LLMs exhibit similar stereotyped approaches to aggregating this variation when asked to predict how members of different social groups interpret the same stimuli. If LLMs can accurately recreate stereotypes in human judgments from first principles in this novel context, this would provide significant evidence that they have internalized processes of stereotype cognition that go beyond simply identifying and retrieving correlations in training data.

\begin{figure*}[htbp]
\centering
\includegraphics[width=0.9\textwidth]{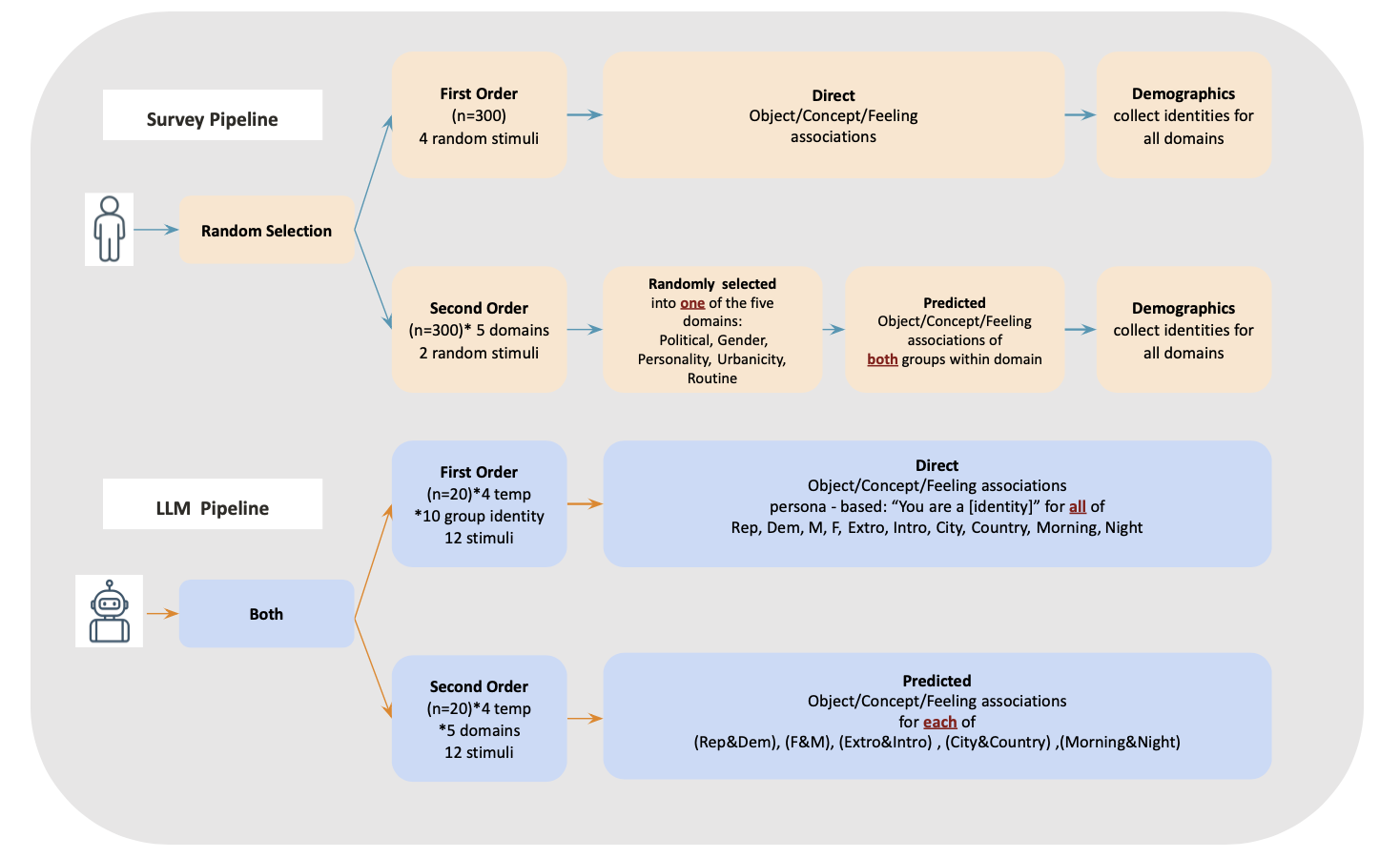}
\caption{Experimental design comparing human and LLM associative processes across first-order and second-order tasks. The survey pipeline (top) shows human participants randomly assigned to either first-order tasks (n=300, providing direct interpretations of 4 random stimuli) or second-order tasks (n=300 per domain, predicting group interpretations for 2 random stimuli for one of the five social domains). The LLM pipeline (bottom) shows both GPT-4o mini and Llama-3.2-11B-Vision-Instruct completing parallel tasks: first-order interpretations using persona-based prompts (``You are a [identity]'') across all group identities and stimuli, and second-order predictions for groups' associations across all five social domains. LLM responses were generated with multiple temperature settings (n=20 per temperature $\times$ 4 temperatures) to capture model variability. Details of the prompts and study design can be found in SI.}
\label{fig:survey_pipeline}
\end{figure*}

\begin{figure*}[htbp]
\centering
\includegraphics[width=0.9\textwidth]{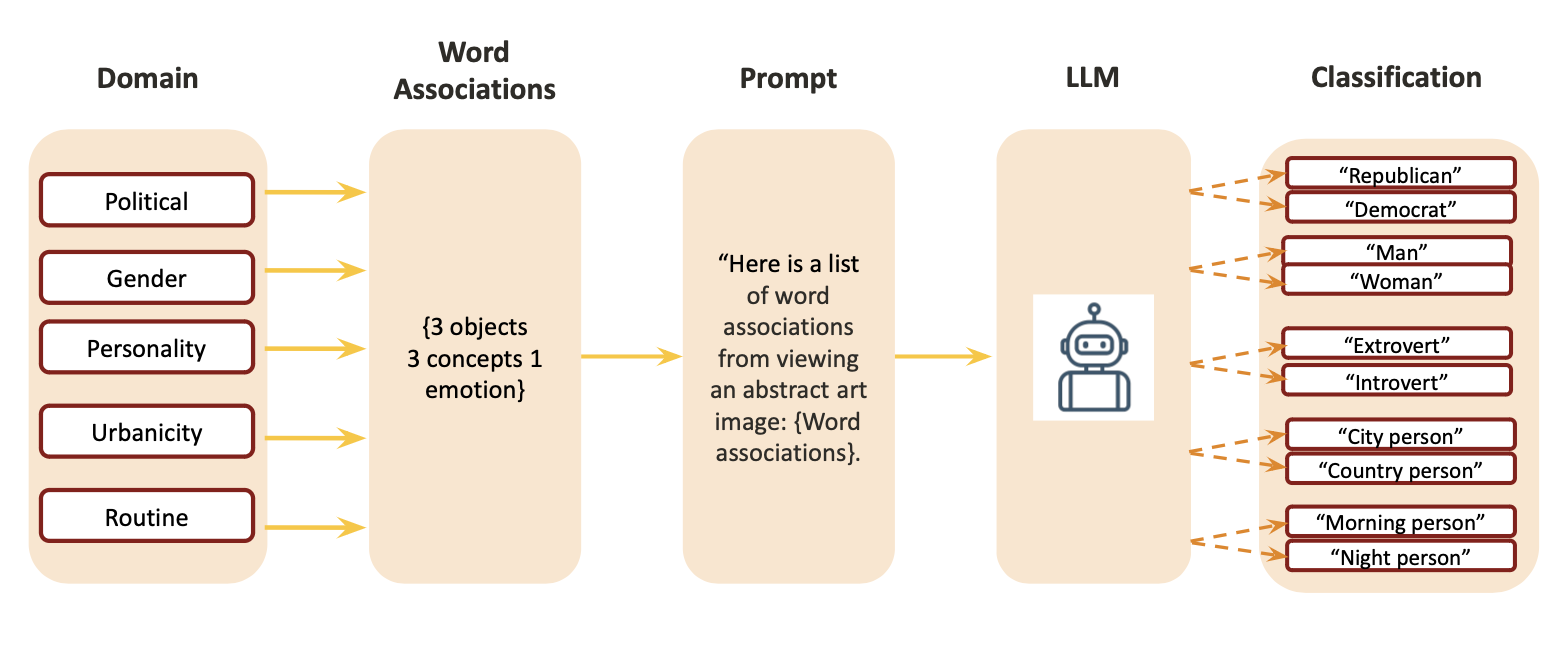}
\caption{Classification methodology for measuring between-group separability of associative patterns. Word associations (3 objects, 3 concepts, 1 emotion) from each social domain by each respondent are fed into a standardized prompt asking LLM classifiers to determine group membership based solely on linguistic patterns. Both GPT-4o mini and Llama-3.2-11B-Vision-Instruct serve as independent classifiers to validate findings across different model architectures. The binary classification task enables quantification of group separability through classification accuracy metrics.}
\label{fig:model_pipeline}
\end{figure*}

Our experimental design is illustrated in Fig.~\ref{fig:survey_pipeline}. We asked participants to self-identify across five domains (gender, partisanship, personality, urbanicity, and routine), chosen because they represent socially established categorical binaries along which people are expected to exhibit and perceive differences in semantic interpretation. We elicited their first-order responses (direct personal interpretations of the stimuli) and second-order responses (predictions about how members of social groups would interpret the stimuli), and replicated this design with two multimodal models (GPT-4o mini and Llama-3.2-11B-Vision-Instruct). We deploy various techniques, including LLM classification, to cluster human and LLM responses within domain by corresponding identities to test whether responses are separable, in either the first or second order, along identity lines (see Fig.~\ref{fig:model_pipeline} for a schematic overview).

We find that human and LLMs differ not only in magnitude but in the structure and semantic content of their stereotypes. Human first-order responses exhibit substantial heterogeneity with only minimal group structure; that is, knowing the demographic information of participants was only weakly predictive of their interpretations, which were highly varied and often idiosyncratic. Yet, when predicting group responses in the second-order, human participants exhibited stereotype exaggeration by moderately amplifying first-order tendencies while still preserving diversity. LLMs' behavior exhibited a qualitatively distinct pattern. LLMs produced homogeneous first-order responses that were similar regardless of the social group they were prompted to identify with; and yet, in the second-order, LLMs generated polarized associations for each domain (gender, partisanship, personality, urbanicity, and lifestyle) that were highly stereotyped, but which departed considerably both in structure and content from both human and LLM responses in the first order. Given the extent to which LLMs' responses in the second-order were untethered from the ground truth first order responses, we refer to this outcome as ``stereotype hallucination.'' Strikingly, we show that LLMs continue to exhibit significant stereotype hallucination even when fine-tuned on the response data of actual participants, suggesting that our findings are not an artifact of minimal or unrepresentative prompting.

\section{Results}
Our experimental design reveals fundamental differences in how humans and large language models navigate the tension between individual associative diversity and group-level categorization when processing ambiguous visual stimuli. We examine two key dimensions: within-group associative diversity (measured by Shannon entropy of word associations) and between-group associative separability (measured by LLMs' classification accuracy of associations by group), comparing responses across first-order (direct association) and second-order (prediction of group association) tasks.

\subsection{Within-Group Associative Diversity Reveals Systematic LLM Homogenization}
\begin{figure*}[htbp]
\centering
\includegraphics[width=0.9\textwidth]{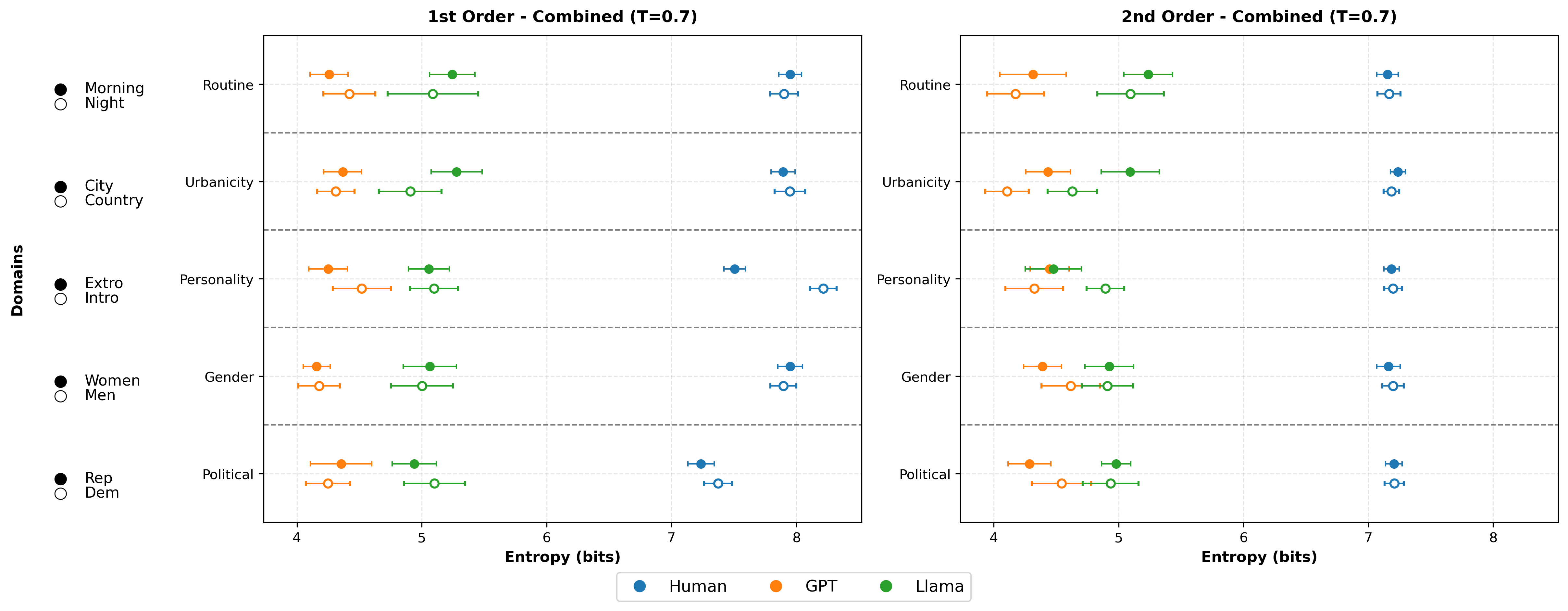}
\caption{Entropy of word association distributions with bootstrap error estimation by order, type of association, domain, and type of respondent. This figure displays the entropy values (displayed on the x-axis) of associations distributions for responses of human participants (blue circles, labeled as Human), GPT-4o mini (orange squares, labeled as GPT) and Llama-3.2-11B-Vision-Instruct (green diamonds, labeled as Llama) for each of the five dichotomized social domain pairs (displayed on the y-axis): political (Republican(Rep)/Democrat(Dem)), gender (women(Wom)/men(Men)), personality (extrovert(Extro)/introvert(Intro)), urbanicity (city(City)/country(Country)), and routine (morning(Morning)/night(Night) person). Each entropy value is measured by calculating the Shannon Entropy of the associations for a given group, order and type of association for a given type of respondent (human, GPT or Llama). 95\% confidence intervals generated using bootstrap are shown. Each group in a domain is colored similarly with slight difference in shape (e.g. solid circle for human Republican response entropy and hollow circle for human Democrat response entropy).}
\label{fig:1}
\end{figure*}

Analysis of associative diversity using Shannon entropy with bootstrap confidence intervals reveals striking differences between human and LLM interpretative patterns (Figure \ref{fig:1}). Across all five social domains, human participants consistently demonstrate significantly higher entropy values than both GPT-4o mini and Llama-3.2-11B-Vision-Instruct for object, concept, and emotion associations. Shannon entropy provides a direct measure of unpredictability in response patterns: higher entropy values indicate greater heterogeneity and diversity of associations within a group, while lower entropy reflects more homogeneous, predictable response patterns. When respondents produce highly varied associations, entropy increases; when they converge on similar associations, entropy decreases.

This pattern persists across experimental conditions. We present here the entropy values for combined associations (object, concept and feeling) and for temperature = 0.7 (the default setting for most language model APIs and is widely used in research and applications) for the two language models.\footnote{Even when LLM temperature parameters were varied (0.3, 0.7, 0.9, 1.0) to maximize output diversity, both GPT-4o mini and Llama maintained substantially lower entropy than human responses. The corresponding figures can be found in the SI.} Notably, while human entropy patterns varied across first-order and second-order tasks depending on social domain (with some domains showing increases rather than decreases in second-order conditions), both LLMs consistently exhibited reduced variability compared to humans in all conditions. This suggests that the compression of associative heterogeneity represents a fundamental characteristic of current LLM architectures rather than a domain-dependent artifact.

\subsection{LLMs Amplify Between Group Separability Through Artificial Stereotype Hallucination}
To quantify the distinctiveness between groups' associative patterns, we employed a direct classification approach that leverages LLMs' ability to detect response patterns associated with different social groups. For each set of word associations (3 objects, 3 concepts and 1 feeling) generated by participants or LLMs, we prompted both GPT-4o mini and Llama-3.2-11B-Vision-Instruct with a consistent binary classification task: ``Here is a list of word associations from viewing an abstract art image: [the set of associations]. Based only on these words, would they most likely come from a [Group A] or [Group B]?'' where Group A and Group B represent the relevant social groups in a domain (e.g., Republican and Democrat) (details in SI). Fig~\ref{fig:model_pipeline} documents this classification methodology.

Classification analysis using both GPT-4o mini and Llama as independent classifiers reveals dramatic differences in how humans and LLMs' responses construct group-based associative patterns (Table~\ref{tab:gpt_classifier}--\ref{tab:llama_classifier}). We employed this dual-classifier approach to validate findings across different LLM architectures and avoid potential classifier-specific biases. Both classifiers show consistent patterns when evaluating responses from humans, GPT-4o mini, and Llama. Higher classification accuracy indicates that the word associations generated by respondents for each group contain more distinctive linguistic signals that reliably differentiate them from the other group's responses, resulting in higher separability between the groups' responses. When accuracy approaches 50\% (random chance for binary classification), it suggests minimal distinguishable differences between groups' associative patterns, indicating low separability. Conversely, accuracy approaching 100\% demonstrates highly separable, group-specific associative patterns.

Across both classifiers and domains, human responses demonstrate moderate increases in classification accuracy from first-order to second-order tasks. In first-order tasks, human classification accuracies cluster remarkably close to random chance across all five social domains (ranging from 46.0--60.0\% depending on classifier and domain). This near-random performance demonstrates that humans do not rely on group-specific associative frameworks when directly interpreting novel, ambiguous stimuli. Individual associative processes appear largely independent of social group membership, supporting theories of cognitive diversity that emphasize personal rather than categorical influences on immediate interpretation. Human second-order tasks show moderate but consistent increases in separability (ranging from 57.4--79.0\% across domains and classifiers). While this indicates that humans do recognize some systematic differences in how social groups might interpret ambiguous stimuli, the moderate accuracy levels suggest these group representations remain nuanced and heterogeneous. The classification accuracies well below perfect performance indicate that human predictions retain substantial variation and resist collapsing into simplified stereotypes, even when explicitly prompted to consider group differences.

In stark contrast, LLM responses exhibit dramatically different patterns that reveal systematic stereotype amplification. In first-order tasks, both GPT-4o mini and Llama show classification accuracies around random chance for most domains, except for personality (up to 83.6\%) accuracy when providing individual persona-based direct interpretations. This suggests that LLMs do not easily default to group typical patterns even with personas provided in first-order tasks. LLM second-order tasks reveal extreme separability that far exceeds human patterns. GPT-4o mini achieves classification accuracies above 90\% for multiple domains under both classifiers, with personality reaching 99.9\% accuracy. Llama shows similarly extreme patterns, with most second-order accuracies exceeding 80\% and several domains approaching or exceeding 95\%. These artificially high separability scores indicate that LLMs construct exaggerated, stereotypical representations where group membership becomes the overwhelming determinant of predicted interpretative patterns, eliminating the nuanced variation that characterizes human cognition. The cross-classifier validation reveals that the extreme separability of LLM responses represents a fundamental characteristic of model architectures rather than an artifact of any single model's classification framework. Both GPT-4o mini and Llama produce responses that are easily distinguished by social group when generating second-order predictions, suggesting systematic stereotype amplification regardless of specific architectural implementations.

\begin{table}[t]
\centering
\caption{Classification accuracy (\%) using GPT-4o mini}
\label{tab:gpt_classifier}
\small
\setlength{\tabcolsep}{6pt}
\begin{tabular}{llccccc}
\toprule
\textbf{Respondent} & \textbf{Order} & \textbf{Political} & \textbf{Gender} & \textbf{Personality} & \textbf{Urbanicity} & \textbf{Routine} \\
\midrule
\multirow{2}{*}{Human} & 1st & 49.3 & 53.8 & 60.0 & 50.8 & 51.9 \\
                       & 2nd & 65.0 & 74.8 & 77.1 & 74.9 & 78.2 \\
\midrule
\multirow{2}{*}{GPT} & 1st & 50.5 & 50.9 & 77.3 & 55.6 & 64.7 \\
                     & 2nd & 90.4 & 84.9 & 99.9 & 95.4 & 97.0 \\
\midrule
\multirow{2}{*}{Llama} & 1st & 64.0 & 59.9 & 80.9 & 72.9 & 65.0 \\
                       & 2nd & 79.3 & 90.4 & 97.2 & 95.0 & 85.7 \\
\bottomrule
\end{tabular}
\end{table}

\begin{table}[t]
\centering
\caption{Classification accuracy (\%) using Llama}
\label{tab:llama_classifier}
\small
\setlength{\tabcolsep}{6pt}
\begin{tabular}{llccccc}
\toprule
\textbf{Respondent} & \textbf{Order} & \textbf{Political} & \textbf{Gender} & \textbf{Personality} & \textbf{Urbanicity} & \textbf{Routine} \\
\midrule
\multirow{2}{*}{Human} & 1st & 46.0 & 48.9 & 54.9 & 49.0 & 50.8 \\
                       & 2nd & 57.4 & 63.8 & 79.0 & 67.0 & 71.8 \\
\midrule
\multirow{2}{*}{GPT} & 1st & 54.0 & 50.7 & 82.8 & 52.7 & 59.1 \\
                     & 2nd & 83.8 & 71.0 & 99.6 & 94.4 & 79.9 \\
\midrule
\multirow{2}{*}{Llama} & 1st & 58.3 & 55.6 & 83.6 & 69.2 & 60.2 \\
                       & 2nd & 71.5 & 79.5 & 95.6 & 94.3 & 80.1 \\
\bottomrule
\end{tabular}
\end{table}

\subsection{Semantic Space Analysis Reveals LLM Construction of Hallucinated Group Prototypes}
\begin{figure*}[htbp]
\centering
\begin{minipage}{0.48\textwidth}
    \centering
    \includegraphics[width=\textwidth]{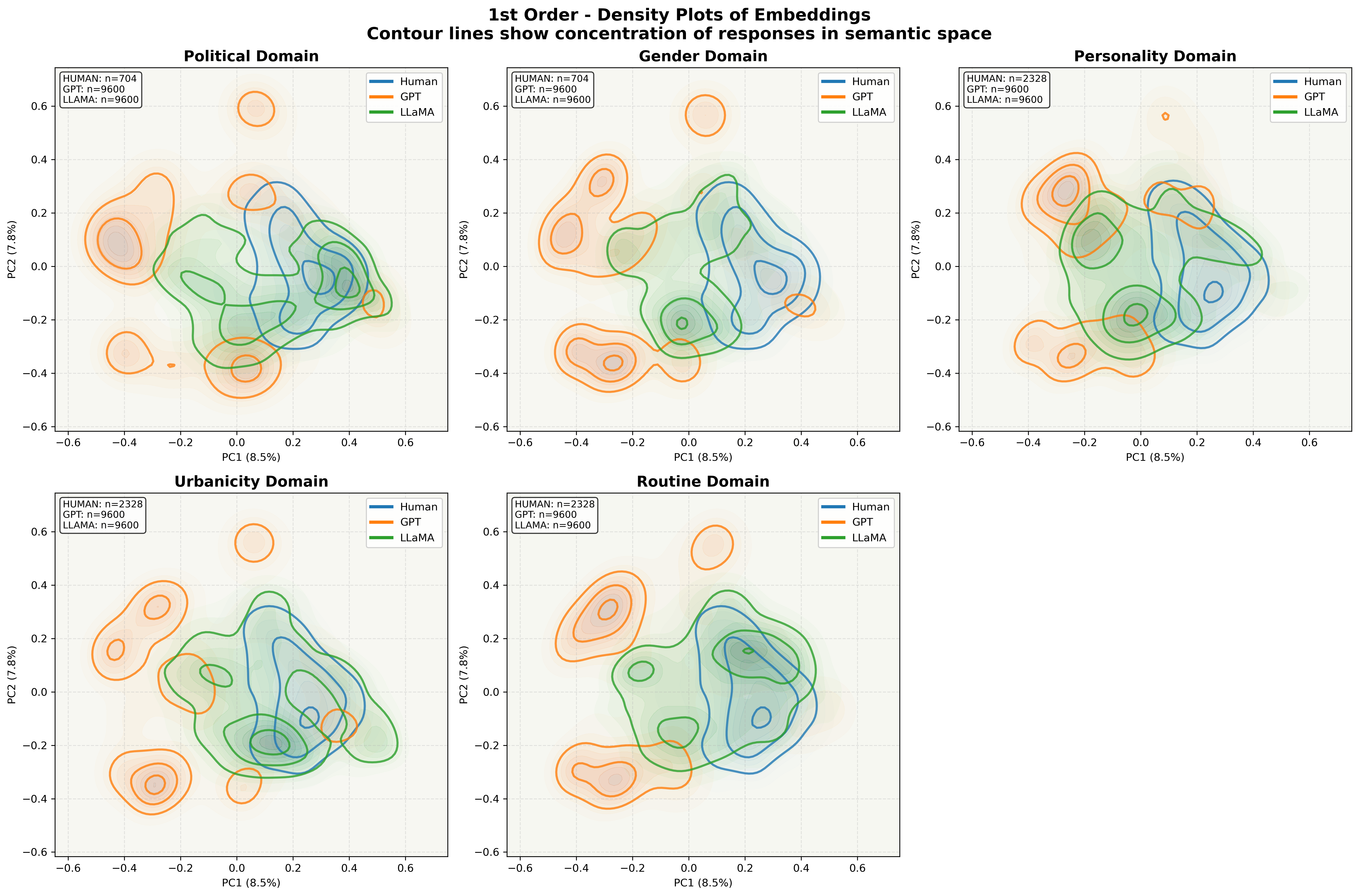}\\
    \textbf{(A)}
\end{minipage}
\hfill
\begin{minipage}{0.48\textwidth}
    \centering
    \includegraphics[width=\textwidth]{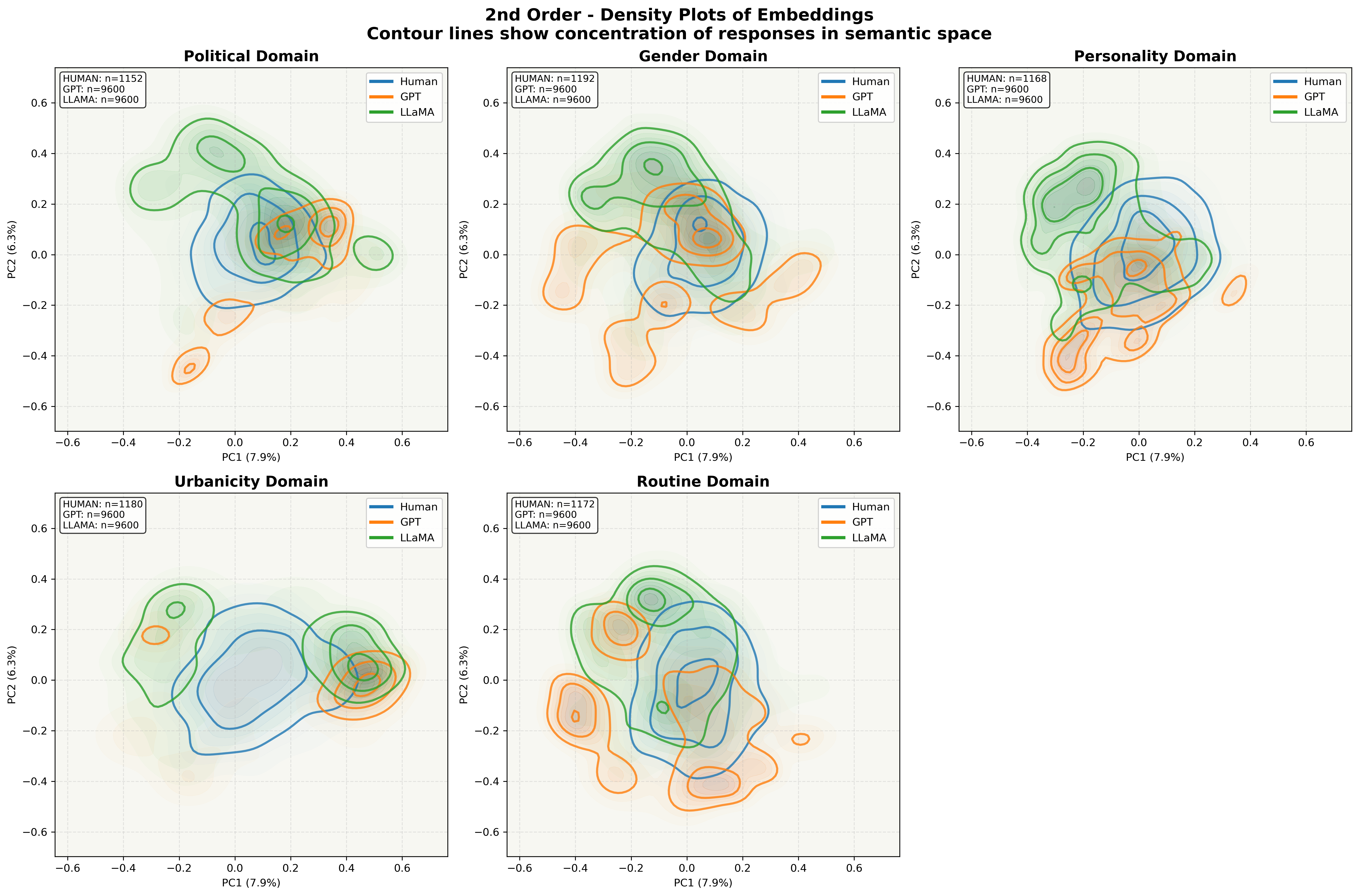}\\
    \textbf{(B)}
\end{minipage}
\caption{\textbf{Semantic space analysis reveals LLM stereotype amplification.}
\textbf{(A)} First-order semantic space distributions show human diversity versus LLM clustering.
Kernel density plots of word association embeddings in two-dimensional PCA space for first-order interpretations across five social domains (political, gender, personality, urbanicity, routine). Each response set (3 objects, 3 concepts, 1 emotion) was encoded using Sentence-BERT (all-MiniLM-L6-v2), then projected using PCA fitted on the full dataset to maintain consistent semantic axes across domains.
Density estimates were computed using Gaussian kernels on a 100$\times$100 grid with contour lines at three levels representing high, medium, and low response concentration. Human responses (blue contours) form broad, overlapping distributions with substantial within-group variation. GPT-4o mini (orange contours) and Llama (green contours) responses show tighter clustering with reduced semantic exploration, even when simulating individual personas.
\textbf{(B)} Second-order distributions reveal LLM construction of hallucinated group stereotypes with extreme semantic separation.}
\label{fig:pca_analysis}
\end{figure*}

While entropy and separability metrics capture distributional properties of associations, principal component analysis of word embeddings in semantic space reveals fundamental differences in the \textit{content} and \textit{topology} of human versus LLM associations (Figure~\ref{fig:pca_analysis}). Moving beyond quantifying variance and group distinctiveness, this analysis examines which semantic regions different types of respondents occupy and how associations are spatially organized within that space. We generated contextual embeddings using Sentence-BERT (all-MiniLM-L6-v2) for each response set (3 objects, 3 concepts, 1 emotion), then projected these into two-dimensional space using PCA fitted on the complete dataset to maintain consistent semantic axes across all domains and conditions. Kernel density estimation with Gaussian kernels visualizes response concentration, with tighter contours indicating concentrated responses and broader contours indicating heterogeneous responses (details in SI Appendix). Density plots of first-order and second-order associations reveal distinct clustering patterns between humans and LLMs across all social domains.

Examining first and second-order semantic distributions (Figure~\ref{fig:pca_analysis}A, Figure~\ref{fig:pca_analysis}B) reveals systematic spatial separation between human and LLM associations. Across all five social domains, human responses (blue contours) form distinct clusters that occupy different regions of semantic space compared to both GPT-4o mini (orange contours) and Llama (green contours). Moreover, human associations consistently form single, continuous clusters across all domains and conditions, creating one contiguous density region with gradually declining concentration from center to periphery, without fragmentation into separate peaks. Llama exhibits intermediate fragmentation, forming moderately separated clusters that maintain some connectivity. GPT-4o mini demonstrates the most extreme fragmentation, often forming multiple tightly separated clusters with minimal bridging density between peaks.

Comparing spatial positions across first-order (Figure~\ref{fig:pca_analysis}A) and second-order (Figure~\ref{fig:pca_analysis}B) tasks reveals qualitatively different patterns. Human distributions show position shifts between task orders but maintain proximity to their first-order locations, representing moderate displacement within the same general semantic neighborhood. GPT 4o-mini exhibit more dramatic spatial relocations to different semantic areas. Llama shows similar but often less extreme relocations, with urbanicity domain associations shifting substantially along both principal components between task orders, forming different clusters of associations. Critically, these relocations differ from the consistent transitions observed in human data. Human shifts maintain approximate orientations and relative positions across domains, suggesting adaptation of existing frameworks. LLM shifts vary in direction and magnitude across domains without apparent systematic relationship to first-order positions, suggesting task-specific rather than framework-adaptive responses.

These semantic space patterns demonstrate that LLM stereotype amplification manifests not only through reduced response diversity (entropy) and increased classification accuracy (separability), but also through the construction of spatially distinct, tightly concentrated semantic clusters that lack the distributed, overlapping character of human associative patterns. The combination of different regional occupation, sharper topological concentration, discontinuous cross-task positioning, and exaggerated within-domain group separation distinguishes LLM from human semantic organization when interpreting ambiguous stimuli.

\subsection{Objections}
\begin{figure*}[htbp]
\centering
\includegraphics[width=0.9\textwidth]{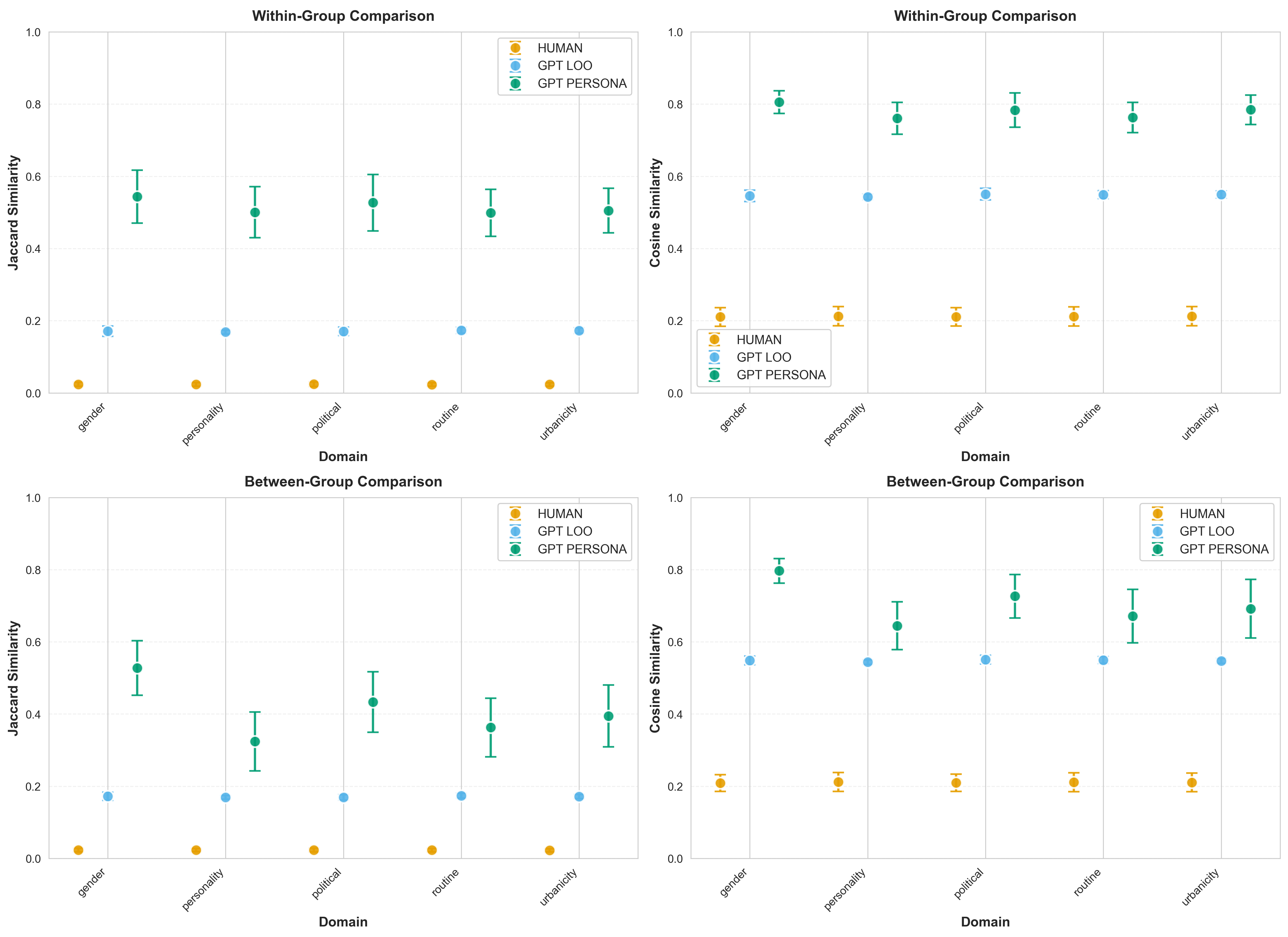}
\caption{\textbf{Similarity analysis across domains for combined associations with fuzzy matching}. Mean similarity scores (points) with 95\% confidence intervals (error bars, computed via standard error) for three data sources (human, orange; GPT-LOO, blue; GPT-Persona, green) across five domains (gender, personality, political, routine, urbanicity). Panels show within-group (top row) and between-group (bottom row) comparisons using Jaccard (left) and Cosine (right) similarity. GPT-Persona shows the highest similarity across domains and metrics, human the lowest, and GPT-LOO intermediate.}
\label{fig:LOO_sim}
\end{figure*}

A critical concern with our experimental design is the asymmetry in comparing approximately 300 individual humans (in each domain) against a single model architecture queried repeatedly with minimal individualization. While human participants bring unique interpretative frameworks shaped by diverse experiences, our LLM prompts provided only group identity (``You are a Republican'') plus temperature variation, raising the possibility that observed homogeneity reflects insufficient individualization rather than fundamental architectural constraints. To address this, we implemented a leave-one-out (LOO) conditioning approach for first-order tasks (details in Materials and Methods and SI Appendix). For each participant, we provided GPT-4o mini with their actual associations to a subset of images and asked it to predict associations for held-out images, giving the model task-relevant, person-specific behavioral data that should maximally enable individual variation capture. Importantly, we avoided conditioning on detailed participant demographics, as this would impose the group-based structure the previous analysis investigates, rather than capturing idiosyncratic patterns underlying interpretation of novel stimuli. Three converging analyses reveal that even optimal conditioning fails to overcome architectural homogenization and stereotype hallucination. Entropy analysis shows GPT-LOO responses remain substantially less diverse than human responses despite exceeding GPT-persona entropy (see Figure S4 in the SI). PCA analysis demonstrates GPT-LOO predictions cluster in tighter and different semantic regions than human responses (see Figure S6 in the SI). Additionally, we conducted image-anchored pairwise similarity analysis to avoid confounding individual differences with stimulus variation. For each image, we formed valid participant pairs among those who viewed that image and computed Jaccard similarity (discrete word overlap) and cosine similarity (Sentence-BERT embeddings) for their associations (three objects, three concepts, one feeling) (details in SI Appendix). We compared mean within-group similarity (both participants from same social group) and between-group similarity (participants from different groups) for each domain with 95\% bootstrap confidence intervals (n=1,000 iterations). Results reveal GPT-LOO pairs exhibit significantly higher within-group similarity than human pairs on both discrete and embedding-based metrics across all domains (Fig~\ref{fig:LOO_sim}). This indicates the model's failure to capture individual heterogeneity, despite conditioning on it. Moreover, GPT-LOO between-group similarity remains nearly as high as within-group similarity. This indicates that the model generates highly similar associations for both groups within a given domain, correctly mirroring the human pattern in which first-order interpretations of novel stimuli show little meaningful group divergences. However, accurately capturing the absence of group structure does not offset the model's more fundamental failure to preserve individual variation. These findings demonstrate that LLM homogenization persists even when models are provided with individuating behavioral context, revealing an architectural constraint rather than a limitation that can be resolved through improved prompting. Models trained on aggregate data privilege central tendencies, and even task-relevant conditioning cannot prevent convergence toward artificial prototypes. Critically, these prototypes differ semantically from actual human responses, suggesting an aggregation process that is disconnected from genuine individual variation characterizing human interpretative processes.

\section{Discussion}
Our results suggest that LLMs engage in free-form interpretive tasks in a qualitatively different way than humans. In contexts where meanings are already stabilized, where stimuli are culturally legible and group-linked associations are entrenched, LLMs can appear to anticipate human responses because human judgments are relatively constrained by shared scripts \cite{goldberg2024interpretation}. We intentionally removed those constraints by using highly ambiguous stimuli chosen precisely because they lack pre-established conventional meanings. In this high-entropy regime, human first-order interpretations remain diverse and only weakly structured by group identity, while human second-order judgments modestly exaggerate whatever weak tendencies exist without collapsing variation into rigid prototypes.

LLMs behave differently. Even when their first-order responses show limited group structure, their second-order outputs produce sharply separable group profiles whose content is untethered from observed first-order patterns. LLMs do not merely exaggerate stereotypes; they hallucinate content that is not grounded in human first-order interpretations. This divergence is perhaps not surprising. LLMs learn about the world exclusively through textual and visual artifacts produced by humans, which disproportionately reflect meanings that have already been filtered, coordinated, and stabilized through social life. They therefore have rich access to the stabilized cultural products of social construction, but limited access to the individual-level perceptual and experiential heterogeneity that generates interpretations under novelty and ambiguity. Humans, by contrast, bring a complex generative cognitive architecture that we still do not fully understand, but that clearly shapes how ambiguous stimuli are interpreted. LLMs' hallucinations persist even when they are conditioned on individual participants' behavioral data, suggesting they fail to emulate this human architecture.

These findings qualify claims that LLMs can serve as ``model organisms'' for the social sciences \cite{park2022social, park2023generative, park2024generative, filippas2024large}. Prior successes in predicting survey responses or partisan reactions \cite{park2024generative, hewitt2024predicting, kozlowski2024silico, kozlowski2025simulating} may depend on domains where meanings and group boundaries are already culturally stabilized, making behavior relatively script-like and therefore learnable as an aggregate pattern from training data. Our use of ambiguous stimuli removes this scaffold and instead tests whether models can generate group-level expectations from heterogeneous first-order interpretations. In this setting, LLMs fail to track the weak structure present in human first-order responses and instead generate prototypical group meanings that are semantically displaced from human interpretations.

This points to a boundary condition on LLM-based social prediction: performance may degrade as contexts become more novel, ambiguous, or culturally unsettled, precisely because human responses become more idiosyncratic and less tightly coupled to demographics. Fine-tuning or conditioning on individual behavioral data may be insufficient if model architectures systematically compress individual variation and default to socially available prototypes when uncertainty is high. If so, LLMs may be least reliable as simulators in the very settings where prediction is most valuable, such as emerging social phenomena and periods before meanings have stabilized. Where the active social construction of meaning from diverse interpretations is at play is where LLMs appear to be least effective in anticipating human behavior.

Several limitations remain. Our entropy comparisons may be sensitive to finite sampling and vocabulary coverage, though separability provides a more direct test of stereotype exaggeration versus hallucination. We studied only two multimodal model families and only visual stimuli; future work should test larger frontier models and other forms of novelty, including unfamiliar auditory stimuli or hypothetical social scenarios. More generally, longitudinal designs that track how initially ambiguous stimuli become culturally legible would offer a direct test of whether LLMs can model the dynamics of meaning stabilization over time as a function of context novelty and complexity.

\section{Materials and Methods}

\subsection{Experimental Design}
We developed a comparative $2 \times 2\times 5$ experimental design to examine how language models and humans interpret an identical set of ambiguous Rorschach blots and abstract art stimuli across first-order (direct interpretation) and second-order (prediction of others' interpretations) tasks across five dichotomized social domains: political affiliation (Republican and Democrat), gender (man and woman), personality (extrovert and introvert), urbanicity (city person and country person) and routine preferences (morning person and night person). We selected 12 visual stimuli comprising 4 Rorschach inkblots and 8 abstract art pieces (images are included in SI) based on three criteria: absence of explicit symbols, sufficient visual complexity to elicit diverse interpretations, and aesthetic neutrality. The deliberate use of abstract, ambiguous stimuli represents a crucial methodological innovation that distinguishes our approach from studies using conventional, socially meaningful stimuli. Because these images lack pre-established meanings or associations, they create an ideal experimental environment for observing how meanings become attached to previously neutral objects through associative processes, without interference from inherited cultural knowledge or learned semantic associations. This design enables isolation of fundamental social construction mechanisms: how individual associative diversity transforms into perceived group patterns, independent of established meanings that would confound results. Fig~\ref{fig:survey_pipeline} shows our overall experimental design.

\subsection{Human Data Collection}
We recruited participants through Prolific, a crowdsourcing platform with established quality controls for academic research. We employed Prolific's representative sampling feature to achieve demographic parity across key variables: age (18-65+), gender (male/female), race/ethnicity (White, Black, Asian, Hispanic, Mixed/Other), and political affiliation (Republican, Democrat, Independent) (details of the sample composition are documented in SI).

We conducted separate surveys for first-order and second-order conditions, recruiting different participants for each survey. In the first-order condition, we recruited participants (n=300) who provided direct associations for four randomly selected stimuli from our set of 12. For each selected stimuli, they provided 3 object associations, 3 concept associations and a feeling which the image elicits from them. They then reported their demographic information, including their relevant social identities across the five social domains of interest at the end of the survey.

For the second-order tasks, we conducted separate surveys for each of the five social domains, recruiting 300 participants for each survey. In these surveys, participants were randomly assigned to complete one of the five surveys corresponding to each domain: political affiliation, gender, personality, urbanicity and routine preferences. To examine stereotype amplification mechanisms, each survey included methodologically motivated false information: ``Based on research examining [domain] (e.g., gender) differences in visual perception, we have found statistically significant variations in associations made by [Group A] (e.g., men) and [Group B] (e.g., women).'' Participants then predicted associations made by \textbf{both} groups within each social domain (e.g., both men and women) using identical association types (3 objects, 3 concepts, a feeling which the image elicits from members of each group) for each group within the domain. This priming enables measurement of how expectations about group differences influence associative attributions, revealing mechanisms of stereotype formation. Similarly, demographic information is collected at the end of surveys. Within each survey, stimuli presentation order and association type order was randomized across participants.

\subsection{Large Language Model Data Collection}
For language model data, we prompted GPT-4o mini and Llama-3.2-11B-Vision-Instruct with carefully controlled inputs to complete both first-order tasks (using persona instructions) and second-order tasks (using prediction instructions) across all 5 social dimensions and 12 stimuli. The prompts used were structurally parallel between human and language model conditions to enable valid comparative analysis.

In first order tasks, because LLMs lack inherent demographic identity, they received structurally parallel instructions with persona priming: ``You are a [Republican/Democrat/Male/Female/etc.]''. They were then prompted to provide 3 object associations, 3 concept associations and 1 feeling. LLMs received identical second-order prompts without persona specifications as human participants, making generalizations based on their learned representations of each social group when interpreting these abstract stimuli. For each domain and each stimulus, the same prompt was used 80 times in total, varying the temperature between 0.3, 0.7, 0.9 and 1 to account for the stochastic nature of the model and to ensure that we captured the variance the models are capable of producing.

\subsubsection{GPT Leave-One-Out Data Collection}
To test whether LLMs can capture individual heterogeneity when provided with person-specific information, we implemented a leave-one-out (LOO) conditioning approach for first-order tasks. For each human participant who viewed four randomly selected images from our 12-stimulus set, we used their actual associations (three objects, three concepts, one feeling) for three images as conditioning context and asked GPT-4o mini to predict what this specific individual would associate with the fourth held-out image. The prompt structure provided the model with explicit individuating information: ``You are a survey participant. You were shown these images [Image 1, Image 2, Image 3] and provided the following associations: [actual participant responses]. You were shown another image \{Image 4 attached\}. What are the three physical objects/abstract concepts which come to mind when you see this image? What is the feeling this image elicits from you?'' We generated responses per participant per held-out image at temperature 0.7, creating a comparable set of LOO predictions for comparison with actual held-out responses, persona-based predictions (using only group identity), and other participants' responses. This design tests whether task-relevant behavioral conditioning enables LLMs to escape group-level stereotypes and approximate individual associative patterns.

\bibliographystyle{unsrtnat}
\bibliography{pnas-sample}

\end{document}